\pdfoutput=1

\documentclass[twocolumn,floatfix]{article}
\usepackage{ulem}  
\usepackage{graphicx}
\usepackage{color}
\usepackage{yhmath}
\usepackage[hidelinks]{hyperref}

\usepackage{xcolor}
\newcommand\comment[1]{\textcolor{red}{#1}}

\begin{document}

\title{The TAP equation: evaluating combinatorial innovation in biocosmology}

\author{Marina Cort\^es,$^{1,2}$ Stuart A.\ Kauffman,$^{3}$ Andrew R.\ Liddle,$^{1,2}$ and Lee Smolin$^1$\\~\\
$^{1}$Perimeter Institute for Theoretical Physics, 31 Caroline Street North, \\ Waterloo, Ontario N2L 2Y5, Canada
\\
$^2$Institute of Astrophysics and Space Sciences, Faculty of Sciences,\\ University of Lisbon, Campo Grande 1769-016 Lisbon, Portugal\\
$^3$Institute for Systems Biology, Seattle, WA 98109, USA
}


\twocolumn[
  \begin{@twocolumnfalse}
    \maketitle
    \begin{abstract}
    \noindent
We investigate solutions to the TAP equation, a phenomenological implementation of the Theory of the Adjacent Possible. Several implementations of TAP are studied, with potential applications in a range of topics including economics, social sciences, environmental change, evolutionary biological systems, and the nature of physical laws. The generic behaviour is an extended plateau followed by a sharp explosive divergence. We find accurate analytic approximations for the blow-up time that we validate against numerical simulations, and explore the properties of the equation in the vicinity of equilibrium between innovation and extinction. A particular variant, the two-scale TAP model, replaces the initial plateau with a phase of exponential growth, a widening of the TAP equation phenomenology that may enable it to be applied in a wider range of contexts.
\end{abstract}
 \vspace*{30pt}
 \end{@twocolumnfalse}
  ]

\thispagestyle{empty}

\tableofcontents

\section{Introduction}

The Theory of the Adjacent Possible (TAP) \cite{TAP} is the notion that the near-future outcomes of some developmental process are limited by the objects that already exist. The objects under study might be physical (molecules, gene sequences, species, etc.) or conceptual (patents, memes, songs, etc.) The TAP equation provides a simple model for the adjacent possible by counting the ways that new objects can be generated from combinations of existing objects. It allows for an efficiency in creation, so that not all possible new objects are realised at each stage.  It is assumed that at each timestep, new objects can be created from any combination of existing objects, while some fraction of existing objects expire (extinction or obsolescence). It is therefore a model of combinatorial innovation \cite{CombInv}.

The number of such objects $M_t$ at time $t$ is then governed by the TAP equation \cite{TAPeqn,TAPeqn2}
\begin{equation}
M_{t+1} = M_t (1-\mu)+ \sum_{i=2}^{M_t} \alpha_i \left( \!\!\! \begin{array}{c} M_t\\ i \end{array}\!\!\! \right) \,.
\label{tap_eq}
\end{equation}
Here the $\alpha_i$ are a set of decreasing constants accounting for the increasing difficulty in linking up larger numbers of elements, the final term is the combinatorial combinations of existing elements, and $0\leq\mu\leq 1$ is the extinction rate of existing objects. As it stands Eq.~(\ref{tap_eq}) is not quite well-defined due to $M_t$ not being constrained to be an integer, but this can be fixed either by analytic interpolation (for instance replacing factorials by Gamma functions) or stochasticity~\cite{TAPeqn2}.

The equation permits dramatically-explosive behaviour, much faster than exponential, because not only do the combinatoric terms rapidly become vast, but so too does the number of terms in the summation. Indeed, typical solutions grow so rapidly as to effectively diverge to infinity in a finite time. Technically they do not diverge at a finite point in time as every term in the summation is finite, despite the number of combinations soon becoming unimaginably large. But this can be considered an artefact of the discretisation and we will see that any analytic approximation to the solution diverges in a finite time, as seen in Ref.~\cite{TAPeqn2}. We refer to this divergence as `blow-up'.

The TAP equation is under active investigation as a possible model for a variety of processes. Koppl et al.~\cite{Koppl2} study three different phenomena: the explosive recent growth of GDP per capita, the rapidly-developing diversity of manufactured goods, and the family tree of US patent applications. In Ref.~\cite{biocosm1} we consider TAP as a model for an evolving state space in the Universe due to emergent laws associated with the appearance of life (see also Ref.~\cite{biocosm2}). Generally, it may be applicable to the collection of phenomena describing the climate indicators --- loosely termed `The Great Acceleration' \cite{greatacc}. It can also be said to model the ongoing impacts of rapid technological development which were first termed `Technological Singularity' in 1993 by NASA's conference proceedings article in Ref.~\cite{vinge} (part of NASA's legacy Cloud-based Data Match-up Service (CDMS) collection). The same phenomenon is also described in the more recent work in Ref.~\cite{Kurzweil}.

In this article we obtain a series of new mathematical results for several variants of the TAP equation, which may be relevant across all its applications.

\section{Approximating solutions to the TAP equation}

We now explore solutions to the TAP equation, both analytically and numerically, for various versions of the equation. The solutions depend on the initial number of elements $M_0$, the extinction rate $\mu$, and the parameters needed to define the $\alpha_i$.

First we make a couple of technical observations. For our results in this section we start the sum in Eq.~(\ref{tap_eq}) at $i=2$, not $i=1$ as in Steel et al.~\cite{TAPeqn2}, in order that the pre-existing elements are not duplicated at each step. However, in an alternative view including the $i=1$ term allows for new items to be created through evolution from a single starting point, and we study this and other TAP variants in  Section~\ref{s:TAPvar}. 

More significantly, we note that Steel et al.'s stochastic implementation is not a discretisation of the TAP equation itself, because it requires a calculational timestep to be small to keep the creation probabilities below unity; it is a stochastic discretization of a continuum approximation to the original equation. Moreover, non-linearity means that even the mean stochastic behaviour will depart from the deterministic evolution. Their numerical analysis also makes the further assumption of a fixed upper limit (usually $i_{\rm max}=4$) in the summation. As we deal directly with the discrete TAP equation, we see some detailed differences in results.

\subsection*{Case 1: constant $\alpha_i$}

When the $\alpha_i$ are equal, $\alpha_i = \alpha$, the sum in Eq.~(\ref{tap_eq}) can be carried out analytically; it is just $\alpha$ times the sum of the corresponding row of Pascal's triangle minus its first two entries, which is $2^{M_t}-M_t-1$. The TAP equation now becomes
\begin{equation}
\label{e:tap_constalpha}
M_{t+1} = M_t (1-\mu) + \alpha \left(2^{M_t} -M_t-1\right) \,,
\end{equation}
whose behaviour can more or less be read off by comparing the relative sizes of the two terms. Except for special values of $\mu$ and $\alpha$ this formula will generate non-integer values of $M_t$ but it can be taken as an analytic continuation to those values (the continuation allowing both for the combinatoric elements and the range of the summation being ill-defined for non-integer $M_t$).

To see the extraordinarily rapid growth this equation implies, we first set $\alpha$ equal to 1 and the extinction $\mu$ to zero. Starting with two elements, $M_0 = 2$, the sequence goes as follows:
\begin{itemize}
\item $t=0$,  $M_0=2$
\item $t=1$, $M_1 = 3$
\item $t=2$, $M_2 = 7$
\item $t=3$, $M_3 = 2^{7} -1 =127$
\item $t=4$, $M_4 = 2^{127} -1 \simeq 10^{38}$
\item $t=5$, $M_5 \simeq 2^{10^{38}} \simeq \exp(10^{38})$ \,.
\end{itemize}
Using $\alpha =1$ and $\mu = 0$ means we are counting {\bf all} the possible combinations that can arise via  the adjacent possible. In later implementations, only a subset are generated.

Only five steps have generated a number much larger than the number of particles in the observable Universe! As the number is shifted into the exponent at each step, a specialist notation for extremely large numbers, known as tetration, described in Section~\ref{s:latetime}, is required to provide an analytic expression.

Blow-up happens as soon as the second term dominates the first, so that each step dramatically changes the total, which then massively enhances the next addition. This is typically when \mbox{$M_t \, 2^{-M_t} < \alpha$.} If this is already satisfied by $M_0$ at $t=0$ blow-up is immediate, otherwise there is initially a plateau of slow growth. As a first estimate  of the length of the plateau we might consider the doubling-time as estimated from the initial growth rate:
\begin{equation}
t_{\rm double} \simeq \frac{M_0}{\alpha  \left(2^{M_0} -M_0 -1\right)} 
\label{e:double}
\end{equation}
steps. This is the time for the cumulated effect of the second term in Eq.~(\ref{tap_eq}) to overcome the first, ignoring extinction and the effects from the growth of $M_t$, the latter assumption clearly rendering this an overestimate.

However we can get a much better and stronger bound on the blow-up time by considering instead the time needed to add just one new item to the set. The insight is that the availability of one new object immediately at least doubles the accessible options, since it can be substituted in place of an item in anything that was previously possible, to create a new object. Hence the rate of growth is at least doubled with the addition of a single new item. The timescale to add one item is a factor $M_0$ smaller than $t_{\rm double}$:
\begin{equation}
t_{\rm add\;one} \simeq \frac{1}{\alpha  \left(2^{M_0} -M_0 -1\right)}  \,.
\end{equation}
Moreover having added one item it must then take no more than half as many steps to add the next, then half again for another, and so on. This series rapidly converges and hence the result should diverge to infinity within $2t_{\rm add\;one}$ steps! All that stops it doing so is that our model consists of discrete steps, at each of which only a finite number can be added, making infinity unreachable. But in any continuum implementation of the equation, for example that of Steel et al.~\cite{TAPeqn2}, divergence in a finite time is inevitable.

We see that the doubling time is a very rapidly decreasing function of time, at least halving with the addition of each new item. By comparison exponential growth, where the doubling time is constant, is very mild indeed!

We now consider the effect of extinction. For any $M_0$ and $\alpha$ combination, there is a critical value of $\mu$ above which extinction dominates growth and ensures $M_t \rightarrow 0$ at late times. Simply comparing the two terms in Eq.~(\ref{tap_eq}), we find 
\begin{equation}
\label{e:mucrit}
\mu_{\rm critical} = \frac{1}{t_{\rm double}} = \frac{\alpha  \left(2^{M_0} -M_0 -1\right)} {M_0} \,.
\end{equation}
Since the extinction rate cannot exceed 1, if this value exceeds unity we are in a regime where the initial growth is so rapid that extinction cannot overwhelm it. Otherwise, for each $(\alpha,M_0)$ pair there is a $\mu_{\rm critical}$ value which separates the high $\mu$ values that lead to total system extinction from the low values where extinction becomes irrelevant at late times. As the evolution in $M_t$ is unstable about the constant value it would have for $\mu = \mu_{\rm critical}$, considerable fine-tuning is required for extinction to have a lasting effect. Without this tuning (or in presence of destabilizing stochastic effects), extinction soon either dominates, or becomes negligible so that the result is not much changed from $\mu = 0$. Hence $\mu$ is not a very crucial parameter.  

\begin{figure}[t]
\centering
\includegraphics[scale=0.9,angle=0]{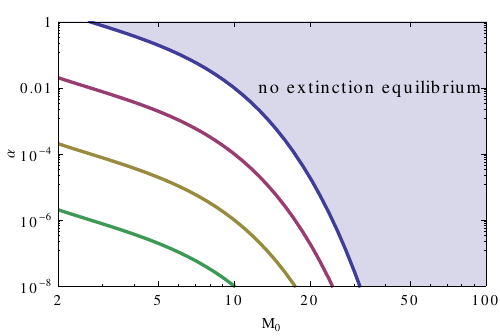}\\
\caption{Critical values of $\mu$ as a function of the parameters $M_0$ and $\alpha$. From top to bottom the contours are $\mu_{\rm critical}$ equal to $1$, $10^{-2}$, $10^{-4}$, $10^{-6}$. The shaded region is where extinction equilibrium is impossible as it would require $\mu > 1$.
\label{extinctionphaseplot}}
\end{figure}

Figure~\ref{extinctionphaseplot} shows Eq.~(\ref{e:mucrit}) contoured in the $(M_0,\alpha)$ plane. In the upper-right region extinction can never overcome growth so there is no extinction equilibrium, whereas otherwise for a given $\mu_{\rm critical}$ the line divides ultimate extinction in the lower left from blow-up in the upper right.

\begin{figure}[t]
\centering
\includegraphics[scale=0.85]{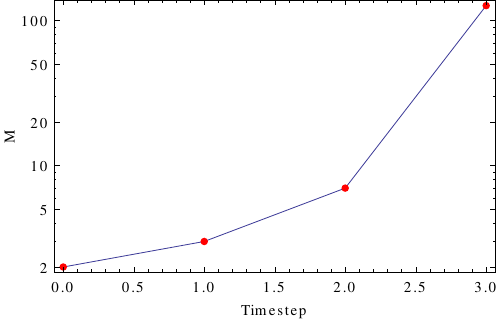}\\
\includegraphics[scale=0.85]{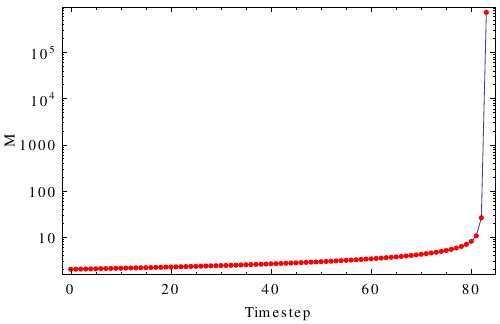}\\
\includegraphics[scale=0.85]{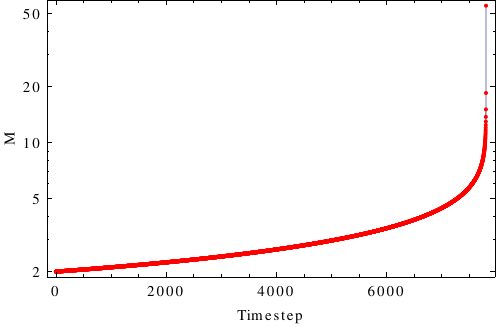}
\caption{Varying $\alpha$ with no extinction and $M_0 = 2$ --- the number of steps to blow-up is inversely proportional to $\alpha$. From top to bottom $\alpha = 1$, $10^{-2}$, and $10^{-4}$. In all figures we omit the last-computed point before computational overflow, to avoid compression of the $y$-axis scale.}
\label{noextinction}
\end{figure}
\begin{figure}[t]
\centering
\includegraphics[scale=0.85]{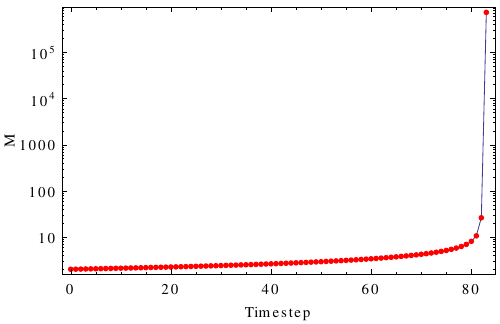}\\
\includegraphics[scale=0.85]{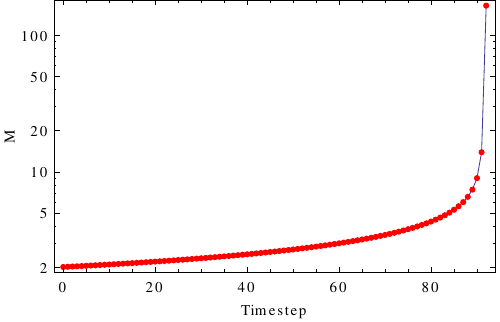}\\
\includegraphics[scale=0.85]{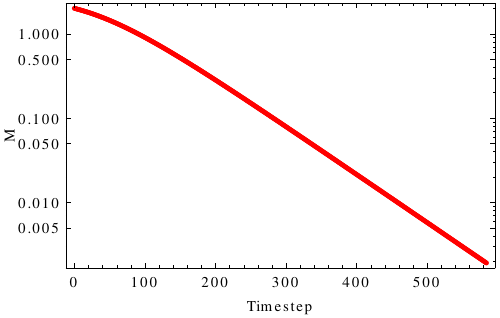}
\caption{Changing extinction at fixed $\alpha = 10^{-2}$ and $M_0 = 2$ --- large enough $\mu$ bends the curve down to zero. From top to bottom $\mu = 0$, $10^{-3}$, and $10^{-2}$.}
\label{constantalpha}
\end{figure}
\begin{figure}[t]
\centering
\includegraphics[scale=0.85]{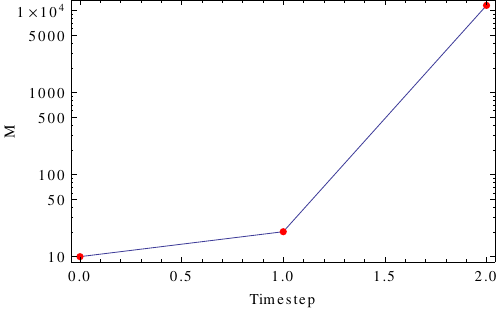}\\
\includegraphics[scale=0.85]{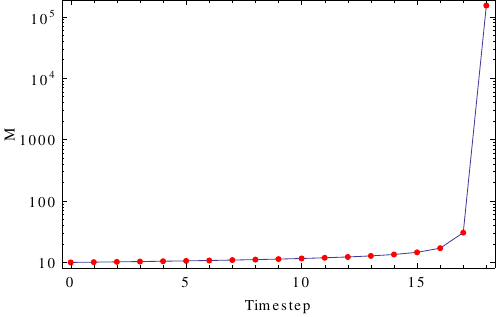}
\caption{As the lower panels of Figure~\ref{noextinction}, but now with $M_0 = 10$.}
\label{biggerM0}
\end{figure}

Numerical results are shown in Figures \ref{noextinction} to \ref{biggerM0} and Table~\ref{t:analyticnumerical} and confirm the validity of the estimates above. The analytic blow-up estimate in the table is taken as $2t_{\rm add\;one}$, which is expected to be an underestimate for small values due to the step discretization, and an overestimate for large values as it underestimates the increase in growth rate (especially for small $M_0$). This is indeed what is seen.

Note that as far as the initial $M_0$ is concerned, the shape of the curve is universal, i.e.\ the evolution onwards from a given value of $M_t$ matches what would be obtained if that value were chosen as the initial condition $M_0$.

The lower part of the table explores some values of the extinction in the vicinity of the critical value for a particular case, showing that even a strong fine-tuning close to the critical value does not substantially prolong the period before blow-up. Steel et al.\ demonstrated that in stochastic models set up close to the instability point, there can be random selection between ultimate blow-up or extinction \cite{TAPeqn2}. Devereaux has recently demonstrated the difficulties of controlling this type of instability in an economic context via policy decisions \cite{Devereaux}.

\begin{table}[t]
\centering
\begin{tabular}{ccccc}
\multicolumn{3}{c}{Values} & Numerical &  Analytic \\ 
$M_0$ & $\alpha$ & $\mu$ &    & \\
\hline
$2$ & $1$ & $0$ & 5 & 2 \\
$2$ & $10^{-2}$ & $0$ & 85 & 200\\
$2$ & $10^{-4}$ & $0$ &7800  & 20000\\
$2$ & $10^{-2}$ & $10^{-3}$ & 94 & 200 \\
$2$ & $10^{-2}$ & $10^{-2}$ & {\rm Never} & {\rm Never}\\
$10$ & $10^{-2}$ & $0$ & 4 & 0.2 \\
$10$ & $10^{-4}$ & $0$ & 20 & 20 \\
\hline
$2$ & $10^{-2}$ & $4\times 10^{-3}$ & 167 & --- \\
$2$ & $10^{-2}$ & $4.9\times 10^{-3}$  & 327 & --- \\
$2$ & $10^{-2}$ & $4.99\times 10^{-3}$ & 504 & --- \\
$2$ & $10^{-2}$ & $4.999\times 10^{-3}$  & 686 & --- \\
 \end{tabular}
\caption{Blow-up times for the constant $\alpha_i$ model, numerical versus analytic estimate. The final blow-up is defined in numerical runs as the first $t$ where $M_t > 10^{10}$, while the analytic estimate is $2 t_{\rm add\; one}$.  Below the line is an investigation of the regime where $\mu$ is in the vicinity of the critical extinction rate $\mu_{\rm critical}$, which is $0.005$ for those $\alpha$ and $M_0$ values, showing that even a highly-tuned initial extinction rate does not much delay blow-up.}
\label{t:analyticnumerical}
\end{table}

%

\subsection*{Case 2: Power-law $\alpha$}

It is reasonable to expect that $\alpha$ is a decreasing function of the index $i$, indicating the difficulty of merging greater numbers of objects to create new ones. A simple parametrisation of this is \cite{TAPeqn2}
\begin{equation}
\alpha_i =\frac{ \alpha}{a^{i-1}} \,,
\end{equation}
such that each additional element to be combined reduces the success rate by a factor of $a$.\footnote{Steel et al.'s simulations in Ref.~\cite{TAPeqn2} use $a=100$, but also cut the sum off at $i_{\rm max} = 4$.}  At least until the vicinity of blow-up when discretisation effects become important, the $\alpha$ parameter is essentially degenerate with the units of the timestepping, provided the extinction is also suitably scaled. For instance if we choose e.g.\ $\alpha = 0.1$ we simply get ten times as many outputs from the code in reaching the same ultimate outcome. It would hence be natural to now set it to one, but for completeness we retain $\alpha$ for now.

The summation can still be done analytically, now equalling $\alpha$ times
\begin{equation}
a \left[ \left( 1+\frac{1}{a} \right)^{M_t} -1 \right] - M_t \,.
\end{equation}
to give
\begin{equation}
\label{e:planal}
M_{t+1} = M_t (1-\mu) + \alpha a \left[\left( 1+\frac{1}{a} \right)^{M_t}  -\frac{M_t}{a}-1\right] \,.
\end{equation}
This is structurally very similar to the original version, which is recovered for $a=1$. Note also that in an expansion of the round brackets for small $1/a$, the two leading terms are cancelled by the other terms in the square bracket, so the leading term from the summation is order $1/a$, showing how its introduction slows the growth rate. 

The crude estimate of the number of steps to double, following Eq.~(\ref{e:double}), is now 
\begin{eqnarray}
t_{\rm double} & \simeq &\frac{M_0}{\alpha a \left[\left(1+1/a \right)^{M_0}-M_0/a - 1\right] } \,; \\
 & \simeq & \frac{2a}{\alpha (M_0-1)} \quad \mbox{for  $a \gg M_0$.}\;\;
\end{eqnarray}
The large $a$ limit shows that the timesteps to double grows linearly with $a$, so reducing the combinatorial efficiency does not qualitatively change the outcome. Once again the critical extinction rate is given by $\mu_{\rm critical} = 1/t_{\rm double}$.

\begin{figure}[t]
\centering
\includegraphics[scale=0.85]{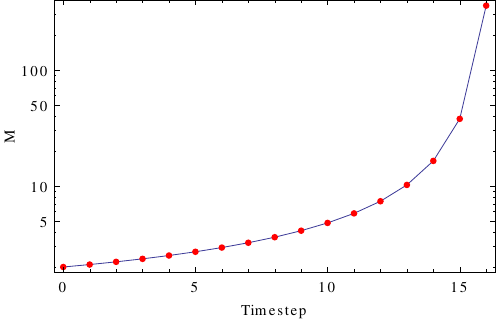}\\
\includegraphics[scale=0.85]{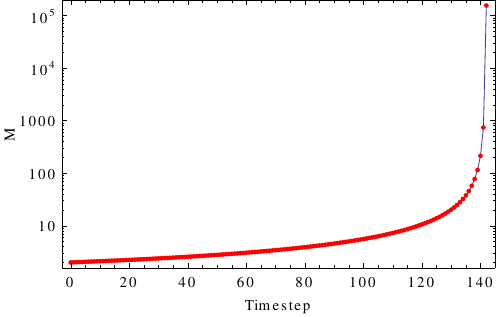}\\
\includegraphics[scale=0.85]{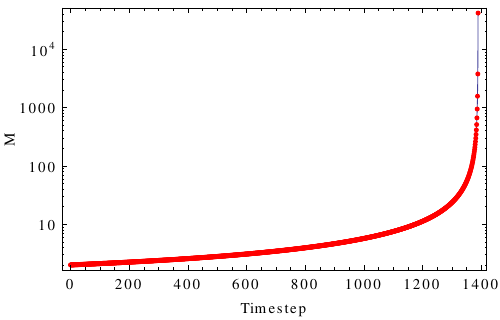}
\caption{Power-law combinatoric suppression at fixed $\alpha = 1$, $M_0 = 2$, and zero extinction. From top to bottom $a = 10$, $100$, and $1000$ (compare also to the top panel of Figure~\ref{noextinction} which is the same with $a=1$). The time to blow-up is linear with $a$  (for $a > M_0$).}
\label{f:plalpha}
\end{figure}

We again examine what happens when we add a single item.  The time required to do so is bounded above by
\begin{equation}
t_{\rm add\;one} \simeq \frac{t_{\rm double}}{M_0}\,.
\end{equation}
While $M_t \ll a$ the growth rate is increased by a factor $(M_t+1)/(M_t-1)$ each time one item is added, but the series sum depends on the starting $M_t$ value and also at some point $M_t$ will become of order $a$ invalidating the above approximation. For large $a$ the series sums to $M_0$. We thus conclude that
\begin{equation}
\label{e:blowupgeneral}
t_{\rm blow-up} \simeq M_0 t_{\rm add\;one}  \simeq \frac{a}{\alpha (M_0-1)} \quad \mbox{for  $a \gg M_0$}\,,
\end{equation}
so that for a given $M_0$, the plateau length grows roughly proportional to $a$. In general a numerical analysis is needed to generate accurate results, which we have carried out. Some evolutions are shown in Figure~\ref{f:plalpha}, and we tabulate some results in Table~\ref{t:analyticnumerical2}.

\begin{table}[t]
\centering
\begin{tabular}{ccccc}
\multicolumn{3}{c}{Values} & Numerical &  Analytic \\ 
$M_0$ & $a$ & $\mu$ &    & \\
\hline
$2$ & $1$ & $0$ & 5 & 1 \\
$2$ & $10$ & $0$ & 18 & 10 \\
$2$ & $100$ & $0$ & 144  & 100 \\
$2$ & $1000$ & $0$ & 1392 & 1000 \\
$6$ & $10000$ & $0$ & 3600 & 2000\\
$10$ & $10$ & $0$ & 5 &  1\\
$10$ & $100$ & $0$ & 25 &  11 \\
$10$ & $1000$ & $0$ & 216 & 111 \\
$100$ & $10000$ & $0$ & 206 & 100 \\
 \end{tabular}
\caption{Blow-up times for the varying $\alpha_i$ model, with $\alpha = 1$. The analytic estimate is Eq.~(\ref{e:blowupgeneral}), whose proportionalities with $a$ and $M_0$ are clearly seen in the numerical results (defined as in Table~\ref{t:analyticnumerical}).}
\label{t:analyticnumerical2}
\end{table}

\section{Late-time TAP}
\label{s:latetime}

The late-time behaviour of the TAP equation is striking for its rate of growth. Taking the simplest case of Eq.~(\ref{e:tap_constalpha}), for sufficiently-large $M_t$ we simply have
\begin{equation}
M_{t+1} \simeq \alpha 2^{M_t} \,.
\end{equation}
At every step the current value is shifted into the exponent, leading to a solution in the form of an exponential tower 
\begin{equation}
M_{t} = 10^{10^{10^{10^{\adots}}}} \,.
\end{equation}
This is the mathematical operation of tetration, for which various mathematical notations exist. The exact number of 10s in the solution, and the power which ultimately appears at the top of the tower, depends on how many steps were taken before the solution entered this limiting phase and at what value.

Despite the extraordinary vastness of these numbers, they are all necessarily finite as the terms in the sums at each step are all finite. This appears to be at odds with our earlier assertion that the doubling time at least halves with each new item created, which implies a divergence in a finite time. The apparent paradox is resolved by a breakdown when the doubling time becomes less than the timestep. While the discrete equation cannot diverge, any continuum approximation to it (or to which it is an approximation) necessarily diverges in a finite time as shown in Ref.~\cite{TAPeqn2}.


\section{More on blow-up}
\label{s:blowup}

We now make some further remarks on blow-up. As already explained, the situation is rather different depending on whether one views the discrete TAP equation as fundamental or as an approximation to a continuous process. In the former case, blow-up is not completely well-defined, as even despite the late-time tetration the evolution does not diverge in a finite number of steps. However, it does not seem reasonable to approximate any process as discrete on a given timescale once it begins evolving rapidly on that timescale. We proposed an operational definition of the time to blow-up using the timescale to add one item and then adding a series of such steps, e.g.\ equation (\ref{e:blowupgeneral}) for the power-law $\alpha_i$ case, which gives an upper limit on when the discrete model will become unjustifiable.

In cases where an analytic continuation of the $M_{t+1}$ formula to real-valued $M(t)$ exists, such as the power-law $\alpha_i$, this can be made more rigorous by taking the continuous limit $dt \rightarrow 0$.\footnote{We thank one of the referees, Mike Steel, for providing this insight.} We then write
\begin{equation}
\frac{dM(t)}{dt} = F[M] \,,
\end{equation}
where $F[M]$ is the analytically-continued function  that gives the evolution, including any specified model parameters such as the $\alpha_i$ [e.g.\ the right-hand side of equation (\ref{e:planal})]. The time to reach infinity from an initial value $M_0$ is then given by the integral
\begin{equation}
t_{{\rm blow-up}} = \int_{M_0}^\infty \frac{dM}{F[M]}\,.
\end{equation}
For a sufficiently-rapidly growing function $F[M]$, which TAP always gives, this will yield a finite value. This provides a rigorously-defined estimate of the effective blow-up time of the discrete model, which can also be interpreted as a time before which the discrete timestepping model must lose viability due to the rate of evolution.

As an example, this formula applied to the fourth row of Table \ref{t:analyticnumerical2} given a blow-up time estimate of 1381, very close to the computed discrete-model blow-up time.

Further investigations of the blow-up are given in this volume by Bellina et al.~\cite{Bellina}, who connect the phenomenon to a number of real-world systems.

\section{TAP equation variants}
\label{s:TAPvar}

\subsection{The `two-scale' TAP equation}

Our main results have taken the combinatorial sum in equation (\ref{tap_eq}) to start at $i=2$, with $i=1$ being omitted to avoid replicating the already-existing objects. However this may not be appropriate in all contexts, because the $i=1$ term can represent significant evolution from a single object. Examples include modifications of an invention motivated only by that invention and not others, or single mutations in DNA caused by radiation damage or transcription error. A new design for a beyblade, for example, may be devised solely to remove a shortcoming of a previous model. Indeed earlier works on the TAP equation typically include the $i=1$ term \cite{TAPeqn,TAPeqn2,Koppl2}, though often accompanied also by an upper cut-off at moderate $i$, perhaps of order 4.

Simply including the $i=1$ term with the same coefficient $\alpha$ does not lead to any substantial differences, for example equation (\ref{e:tap_constalpha}) just loses the $M_t$ from the final bracket leading to slightly faster growth. But for some highly-complex items, mutations of single sources is likely to be the dominant evolutionary driver; e.g.\ dogs evolve through genetic mutations and through breeding with other dogs, not because their genome is successfully merged with that of a coffee bean that they happened to chew. This general idea motivates a study of TAP in the form
\begin{equation}
M_{t+1} = M_t (1-\mu)+ \alpha_1 M_t + \sum_{i=2}^{M_t} \alpha \left( \!\!\! \begin{array}{c} M_t\\ i \end{array}\!\!\! \right) \,,
\label{tap_eq_2scale}
\end{equation}
where $\alpha_1$ is the rate of evolution from individual objects and $\alpha \ll \alpha_1$ is the multi-object combination rate from before,  now highly suppressed relative to the evolution term. We see immediately that the evolution rate is perfectly degenerate with the extinction $\mu$, hence all results go through as before, except that we are now motivated to consider $\alpha_1 > \mu$ that corresponds to the previously impossible case of anti-extinction.

\begin{figure}[t]
\centering
\includegraphics[scale=0.9,angle=0]{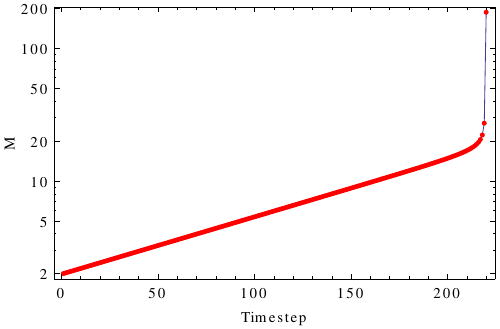}\\
\caption{Evolution with $\alpha_1 = 0.01$, $\alpha = 10^{-6}$ and no extinction (or more generally any values with $\alpha_1 - \mu = 0.01$ would give the same result). We see an initial exponential phase that then gets overtaken by the TAP blow-up.}
\label{f:2scale}
\end{figure}

Within this regime, rather than a plateau followed by an explosion, we expect an initial exponential phase [as $(M_{t + 1} - M_t) \propto M_t$], until the blow-up is initiated by $M_t$ reaching a sufficient value to overcome the suppression of the combinatoric term by the small $\alpha$. Figure~\ref{f:2scale} confirms this. It is notable that the exponential portion of the curve is very accurately so, and hence gives essentially no warning of the impending sharp transition to the TAP blow-up. 

The timescale to blow-up is readily estimated by the techniques used earlier in this article.\footnote{For simplicity we have assumed a single rate $\alpha _i = \alpha$ for $i \geq 2$. We also analysed with a power-law as earlier, but this cannot extend the time to blow-up significantly since $M_t$ is already rising exponentially and hence the combinatoric sum as a double exponential.} Using the same guideline of balancing the contributions to the evolution, here giving $M_t 2^{-M_t} \simeq \alpha/\alpha_1$, combined with the exponential growth predicted by the leading term predicts a blow-up at $M_t \simeq 17$ and at around timestep 210 for the parameters of Figure~\ref{f:2scale}, in excellent agreement with the full calculation.

\subsection{The differential TAP equation}
\label{s:diffTAP}

The TAP equation with $\alpha = 1$ and no extinction seeks to count {\bf all} adjacent possibilities. However, the equation regenerates the same items over and over at each step, so that, for instance, an object first created at step 2 is then duplicated at each subsequent step from the same constituents. We can correct for this simply by subtracting the number of items created at the previous step. Taking care to retain the original object set $M_0$ the equation now reads
\begin{equation}
M_{t+1} = M_0+ \sum_{i=2}^{M_t} \left( \!\!\! \begin{array}{c} M_t\\ i \end{array}\!\!\! \right) \,.
\label{difftap_eq}
\end{equation}
so that the first term is now always $M_0$ rather than $M_t$. This only slightly moderates the growth compared to the original TAP equation, the sequence for $M_0 = 2$ now being $\{2,3,6,59,6\times 10^{17},...\}$,
and hence is not a very interesting alternative.

There's no simple way to carry out the same process for $\alpha_i <1$ as we would then need to track which individual objects were actually created at each step to excise duplicates.

\subsection{The logistic TAP equation}
\label{s:logTAP}

We mention one further idea which we reserve for future investigation: the question of whether there is a logistic TAP equation. In population dynamics, the logistic mapping arises when the birth rate is moderated by making the parents compete for resources, most simply modelled by adding a $-M^2$ term \cite{May}. This model already shows period doubling and beautiful deterministic chaos. 

A TAP equivalent would be to consider that the ability to generate and test inventions is limited by the number of inventors. The goal is to translate that concept into an $M$-dependent growth suppression term in TAP, to see if it can moderate the later stages of the explosion.

\section{Conclusions}

In this section we will examine the TAP equation's distinctive `hockey-stick' property. We could describe that property as a long term period of calm dynamics suddenly interrupted by a `blow-up' evolution: {\it it was going along smoothly, then all of a sudden it TAPs}.

The TAP equation is in fact a family of equations, whose properties can to some extent be tuned to best fit the circumstances in which it is being applied. To fully specify it requires us to choose the range of the index $i$ of the combinatoric term, the extinction rate $\mu$, and either individual $\alpha_i$ for each term in the sum or a parametrized functional form that they can be derived from. For most versions the generic behaviour is a flat plateau, followed after some time by a sharp super-exponential blow-up \cite{TAPeqn,TAPeqn2}. This `hockey-stick' shape appears to capture the observed form of a wide variety of phenomena, as discussed for instance in Ref.~\cite{Koppl2}. We have provided accurate analytical formulae for the time to blow-up for several versions of the TAP equation, and verified their accuracy via numerical simulations.  Our results complement and extend those of Steel et al.~\cite{TAPeqn2} who analysed a stochastic variant of TAP.

We have also explored a new regime for TAP in a two-scale variant, so-named because the timescale for significant evolution from a single object is much shorter than the timescale to form new objects via combinations. In this case the initial plateau is replaced by a phase of exponential growth, which is still followed by a blow-up once the combinatoric timescale is reached.

There are hints in the literature that TAP behaviour might describe the ongoing environmental catastrophes \cite{greatacc}.  Ultimately economic development is what leads to most forms of overexploitation of the planet, hence if TAP underlies economic development, as proposed in Refs.~\cite{TAPeqn,Koppl2}, it must also have a role to play in interpreting environmental and climatic indicators. If this is correct the implications are profound, and deeply concerning. Our findings show that the transition to the TAP blow-up is sudden, explosive, and is not foreshadowed by any features of the curves until its onset. It is hence essentially impossible to predict in advance the timing of a TAP blow-up, until it is already underway. This statement might prove critical for the correct understanding of the environmental situation we find ourselves in. Not only that, but TAP demonstrates potential grave implications for the sustainability of our species in the short term (meaning the extinction of the human species in the time range of order decades), if it is indeed the correct description of climatic phenomena.

\subsection*{Acknowledgments}

We thank Abby Devereaux, Stephen Guerin, Wim Hordijk, Roger Koppl, Lars Larsen, Andrew Sheng, Kilian Srowik, and Mike Steel for discussions.

This research was supported in part by Perimeter Institute for Theoretical Physics. Research at Perimeter Institute is supported by the Government of Canada through Industry Canada and by the Province of Ontario through the Ministry of Research and Innovation. This research was also partly supported by grants from NSERC and FQXi. This work was supported by the Funda\c{c}\~{a}o para a Ci\^encia e a Tecnologia (FCT) through the research grants UIDB/04434/2020 and UIDP/04434/2020. M.C.\ acknowledges support from the FCT through grant SFRH/BPD/111010/2015 and the Investigador FCT Contract No.\ CEECIND/02581/2018 and POPH/FSE (EC). A.R.L.\ acknowledges support from the FCT through the Investigador FCT Contract No.\ CEECIND/02854/2017 and POPH/FSE (EC). M.C.\ and A.R.L.\ are supported by the FCT through the research project EXPL/FIS-AST/1418/2021. We are especially thankful to the John Templeton Foundation for their generous support of this project.

\end{document}